\documentclass[aps,pra,reprint,superscriptaddress,letterpaper]{revtex4-1}

\usepackage{amsmath,amsthm,amsfonts,amssymb,bm}
\usepackage{times}
\usepackage[colorlinks={true},dvipdfm]{hyperref}
\hypersetup{citecolor={blue}, filecolor={blue}, linkcolor={blue}, urlcolor={blue}}
\usepackage{float}
\usepackage{epsf}
\usepackage{color}
\usepackage{verbatim}
\usepackage{enumerate}
\usepackage{multirow}
\usepackage{anysize}
\usepackage{graphicx}
\usepackage{comment}
\usepackage{bm}

\begin{document}

\title{Searching for quantum optimal controls under severe constraints}
\author{Gregory Riviello}
\affiliation{Department of Chemistry, Princeton University, Princeton, New Jersey 08544, USA}
\author{Katharine Moore Tibbetts}
\affiliation{Department of Chemistry, Princeton University, Princeton, New Jersey 08544, USA}
\affiliation{Department of Chemistry, Temple University, Philadelphia, PA 19122, USA}
\author{Constantin Brif}
\affiliation{Department of Scalable \& Secure Systems Research, Sandia National Laboratories, Livermore, CA 94550, USA}
\author{Ruixing Long}
\affiliation{Department of Chemistry, Princeton University, Princeton, New Jersey 08544, USA}
\author{Re-Bing Wu}
\affiliation{Department of Automation, Tsinghua University and Center for Quantum Information Science and Technology, TNlist, Beijing, 100084, China}
\author{Tak-San Ho}
\affiliation{Department of Chemistry, Princeton University, Princeton, New Jersey 08544, USA}
\author{Herschel Rabitz}
\affiliation{Department of Chemistry, Princeton University, Princeton, New Jersey 08544, USA}

\begin{abstract}
The success of quantum optimal control for both experimental and theoretical objectives is connected to the topology of the corresponding control landscapes, which are free from local traps if three conditions are met: (1) the quantum system is controllable, (2) the Jacobian of the map from the control field to the evolution operator is of full rank, and (3) there are no constraints on the control field. This paper investigates how the violation of assumption (3) affects gradient searches for globally optimal control fields. The satisfaction of assumptions (1) and (2) ensures that the control landscape lacks fundamental traps, but certain control constraints can still introduce artificial traps. Proper management of these constraints is an issue of great practical importance for numerical simulations as well as optimization in the laboratory. Using optimal control simulations, we show that constraints on quantities such as the number of control variables, the control duration, and the field strength are potentially severe enough to prevent successful optimization of the objective. For each such constraint, we show that exceeding quantifiable limits can prevent gradient searches from reaching a globally optimal solution. These results demonstrate that careful choice of relevant control parameters helps to eliminate artificial traps and facilitate successful optimization.
\end{abstract}

\maketitle

\section{Introduction}
\label{sec:intro}

Applications of quantum control in the laboratory have grown dramatically over the past fifteen years \cite{Rabitz2000, LevisRabitz2002, Goswami2003, BrixnerGerber2003, Nuernberger2007, DantusLozovoy2004, Brif2010NJP, WollenhauptBaumert2011, Brif2012ACP}. Successful optimal control experiments (OCEs) have included selective control of molecular vibrational \cite{HornungMeierMotzkus2000, WeinachtBartels2001CPL, BartelsWeinacht2002PRL, KonradiSingh2006JPPA, KonradiScaria2007, ScariaKonradi2008, StrasfeldShim2007, StrasfeldMiddleton2009} and electronic states \cite{BardeenYakovlevWilson1997, BrixnerDamrauer2001, Nahmias2005, Prokhorenko2005, BonacinaWolf2007, KurodaKleiman2009, vanderWalleHerek2009, RothGuyonRoslund2009, RoslundRothGuyon2011, WeiseLindinger2011}, preservation of quantum coherence \cite{Branderhorst2008, BiercukBollinger2009}, control of photoisomerization reactions \cite{VogtKrampert2005, DietzekYartsev2006, DietzekYartsev2007, ProkhorenkoNagy2006, VogtNuernberger2006CPL, GreenfieldMcGrane2009}, selective manipulation of chemical bonds \cite{Assion1998, BergtBrixner1999, Levis2001, VajdaBartelt2001, Daniel2003, Plenge2011, NuernbergerWolpert2010, NuernbergerWolpert2012, MooreXing2013PCCP}, high-harmonic generation and coherent manipulation of the resulting soft X-rays \cite{Bartels2000, Bartels2001, Bartels2004, Reitze2004, PfeiferKemmer2005, PfeiferSpielmann2006, Winterfeldt2008}, and control of energy flow in biomolecular complexes \cite{Herek2002, WohllebenBuckup2003, BuckupLebold2006, SavolainenHerek2008}. Optimal control theory (OCT) \cite{RabitzZhu2000, DAlessandro2007, WerschnikGross2007, BalintKurti2008, Brif2010NJP, Brif2012ACP} has facilitated an improved understanding of coherently controlled quantum phenomena such as electron density transfer \cite{Kammerlander2011, Castro2012}, electron ring currents in molecules \cite{KannoHoki2007}, molecular photodissociation \cite{Kosloff1989, ShiRabitz1991, Gross1991, Gross1992, NakagamiOhtsuki2002, Krieger2011}, photoisomerization \cite{OhtsukiOhara2003, ArtamonovHo2004CP, ArtamonovHo2006CP, ArtamonovHo2006JCP, KurosakiArtamonov2009} and photodesorption \cite{NakagamiOhtsukiFujimura2002CPL}, strong-field ionization \cite{RasanenMadsen2012}, quantum information processing \cite{TeschVivieRiedle2002, VivieRiedleTroppmann2007, PalaoKosloff2002, PalaoKosloff2003, KhanejaReiss2005, SchulteSporl2005, DominyRabitz2008JPA, Schirmer2009JMO, Nebendahl2009, TsaiChenGoan2009, ZhuFriedrich2013, Hohenester2006, GraceBrif2007JPB, GraceBrif2007JMO, GraceDominy2010NJP, FloetherSchirmer2012, MontangeroCalarcoFazio2007PRL, SafaeiMontangero2009, WeninPotz2008PRA, WeninPotz2008PRB, WeninRoloffPotz2009, RoloffWeninPotz2009JCE, RoloffWeninPotz2009JCTN, MerkelDeutsch2009, DeutschJessen2010, MischuckDeutsch2012, Rebentrost2009PRL, RebentrostWilhelm2009, MotzoiGambetta2009PRL, SchulteHerbruggenSporl2011, KosutGrace2013, HockerBrif2014}, energy transfer in photosynthetic complexes \cite{BruggemannMay2004CPL, BruggemannMay2004JPCB, BruggemannMay2006, BruggemannMay2007, Caruso2012}, transport of Bose-Einstein condensates \cite{HohenesterRekdal2007, HohenesterGrond2009, GrondHohenester2009}, and transport of atoms in optical lattices \cite{ChiaraCalarco2008, DoriaCalarco2011, MischuckDeutsch2010}. 

In general, the goal of OCE and OCT is to find a control field $\varepsilon(t)$ that produces the global maximum or minimum value of an \textit{objective functional} $J = J[\varepsilon(t)]$. This functional represents quantum control objectives such as the probability of a transition between two pure states, the expectation value of an observable, or the distance between a target unitary transformation and the time-evolution operator \cite{Brif2012ACP}. The \emph{quantum control landscape} defined by this functional dependence has been depicted in experimental studies for various control problems \cite{VogtNuernberger2006PRA, RoslundRabitz2006, MarquetandNuernberger2007, FormWhitaker2008, RoslundRabitz2009PRA_2, RuetzelStolzenberger2010, SchneiderWollenhaupt2011, MooreXing2013JCP}, and its favorable topology \cite{ChakrabartiRabitz2007, Brif2012ACP} has been correlated \cite{RabitzHsiehRosenthal2004, HoRabitz2006JPPA, MooreHsiehRabitz2008JCP} to the success of OCEs and OCT simulations. Specifically, it has been shown \cite{RabitzHsiehRosenthal2005PRA, HsiehRabitz2008PRA, HoDominyRabitz2009PRA, HsiehWuRabitzLidar2010, RabitzHoHsieh2006PRA, HoRabitz2006JPPA, RabitzHsiehRosenthal2006JCP, WuRabitzHsieh2008JPA, HsiehWuRabitz2009JCP} that the landscapes for $N$-level closed quantum systems lack local optima if three conditions are satisfied: (1) the quantum system is controllable, i.e., any given unitary evolution can be generated by some control field in finite time; (2) the Jacobian of the map from the control field $\varepsilon(t)$ to the final-time evolution operator $U(T,0)$ is of full rank; (3) the control field is unconstrained. We discuss these conditions in more detail in Sec.~\ref{sec:back}. Local optima can potentially trap a gradient search, so their absence from the control landscape facilitates identification of a globally optimal control field. Although the control landscape topology \cite{Pechen2008JPA, WuPechenRabitz2008JMP, WuRabitz2012JPA, PechenBrif2010PRA} and optimization search effort \cite{OzaPechen2009JPA} for open quantum systems have been studied, we do not consider issues related to open-system control in this work.

In this paper, we assume that conditions (1) and (2) have been met but that condition (3) is subject to violation. We consider several types of control constraints: the representation of the control field, the number of control variables, the duration of the control pulse, the field strength, and several parameters of the search algorithm. The nature of the search algorithm falls under assumption (3), as it can artificially limit access to desired controls in some circumstances. For each of these constraints, we perform a large number of numerical OCT searches on a variety of closed, finite-level quantum systems, accruing statistical evidence of each constraint's effect on the gradient optimization of various quantum objectives. These numerical studies make it possible to quantify the limits beyond which the severity of a constraint leads to the emergence of artificial local traps on the control landscape and hinders the achievement of a globally optimal solution.  In most cases, we identify two key values of the constrained parameter: one beyond which at least one search fails (indicating the emergence of traps on the control landscape), and one beyond which all searches fail (suggesting that the global optimum is unreachable).

The remainder of this paper is organized as follows: Section~\ref{sec:back} discusses the classification of landscape critical points and the theoretical underpinnings of conditions (1) -- (3). Section~\ref{sec:methods} describes the control objectives used in this paper, the topology of the corresponding landscapes, and the numerical methods used to optimize them. In Sec.~\ref{sec:resources}, we examine how searches for globally optimal solutions are influenced by severe constraints on the control field, which may prevent successful optimization. Our conclusions are summarized in Sec.~\ref{sec:concl}.

\section{Background}
\label{sec:back}

The control problems discussed in this paper are defined as closed $N$-level quantum systems whose Hamiltonians have the form
\begin{equation}
\label{eq:ham-1}
H(t) = H_0 + \sum_{k=1}^K H_k \varepsilon_k (t) ,
\end{equation}
which includes a field-free term $H_0$ and $K$ Hermitian operators $\{ H_k \}$ that represent the coupling between $K$ control fields $\{\varepsilon_k (t) \}$ and the system. Each field is a real-valued function of time defined on the interval $[0,T]$. In the Schr\"{o}dinger picture, the state of the system at a time $t$ is described by the state vector $|\psi(t)\rangle = U(t) |\psi_0\rangle$ or, for mixed states, by the density matrix $\rho(t) = U(t) \rho_0 U^{\dagger}(t)$. Here, $|\psi_0\rangle \equiv |\psi(0)\rangle$ is the initial state vector, $\rho_0 \equiv \rho(0)$ is the initial density matrix, and $U(t) \equiv U(t,0)$ is the time-evolution operator or propagator. $U(t)$ satisfies the Schr\"{o}dinger equation:
\begin{equation}
\label{eq:schro}
i \hbar \frac{d}{dt} U(t) = H(t) U(t) , \ \ \ U(0) = \mathbb{I} ,
\end{equation}
where $\mathbb{I}$ is the $N$-dimensional identity operator. 

A quantum system that obeys the Schr\"{o}dinger equation is \textit{evolution-operator controllable} \cite{DAlessandro2007, Brif2012ACP} if for any unitary operator $W$ there exists a set of controls $\{\varepsilon_k (t) \}$ such that $W$ is the solution to Eq.~\eqref{eq:schro} at some finite time. For a system governed by the Hamiltonian of form \eqref{eq:ham-1}, the necessary and sufficient condition for evolution-operator controllability is that the Lie algebra generated by the set of operators $(i/\hbar)\{ H_0, H_1, \ldots ,H_K \}$ be u($N$) [or su($N$) if the Hamiltonian has zero trace] \cite{Ramakrishna1995, SchirmerSolomon2002JPA_1, SchirmerSolomon2002JPA_2, Albertini2003, Altafini2009}. A previous work \cite{WuHsiehRabitz2011PRA} has examined the loss of controllability and the resulting local traps on the control landscape, but in this paper, we only study systems that are assumed to satisfy this controllability criterion. We consider control problems that employ one control field $\varepsilon(t)$ except when specifically noted otherwise.  In this limiting case, Eq.~\eqref{eq:ham-1} simplifies to the Hamiltonian of the form 
\begin{equation}
\label{eq:ham-2}
H(t)  = H_0 - \mu \varepsilon(t),
\end{equation}
which arises in the electric dipole approximation; the \emph{dipole operator} $\mu$ couples the system to the field. In the remainder of this section and in Sec.~\ref{sec:methods}, we assume that the Hamiltonian has the form in Eq.~\eqref{eq:ham-2}.  It is straightforward to generalize the analysis to Hamiltonians of the form in Eq.~\eqref{eq:ham-1}.

\textit{Critical points} of a quantum control landscape are the set of control fields at which the first-order functional derivative of the objective $J$ with respect to the control field is zero:
\begin{equation}
\label{eq:crit}
\frac{\delta J}{\delta \varepsilon(t)} = 0 , \ \ \ \forall t \in [0,T] .
\end{equation}
The topology of the control landscape is determined by the classification of critical points according to the properties of the higher-order functional derivatives of $J$; critical points can be characterized as local optima, global optima, and saddles \cite{ChakrabartiRabitz2007, Brif2012ACP}.  The landscape topology has practical significance for quantum control optimizations, since local optima may trap gradient searches and can even affect the efficiency of genetic algorithms \cite{DigalakisMargaritis2001}. When the landscape lacks local traps, on the other hand, several OCT studies consisting of thousands of numerical simulations have shown that gradient searches can quickly locate globally optimal controls \cite{MooreHsiehRabitz2008JCP, MooreChakrabarti2011, MooreRabitz2011, RivielloRabitz-prep}.  In the laboratory, a gradient algorithm \cite{RoslundRabitz2009PRA_1} and a derandomized evolution strategy \cite{RoslundShir2009PRA} have been successfully employed to make OCEs more efficient.

The landscape analysis also draws the important distinction between regular and singular critical points \cite{BonnardChyba2003, ChakrabartiRabitz2007, Brif2010NJP, Brif2012ACP}. Further partitioning the functional relationship between the objective $J$ and the control field $\varepsilon(t)$, we can represent $J$ as a function of the final-time evolution operator $U_T \equiv U(T)$, and $U_T$ in turn as a functional of the control field; i.e., $J = J(U_T)$ and $U_T = U_T [\varepsilon(t)]$. We then use the chain rule to rewrite Eq.~\eqref{eq:crit} as: 
\begin{equation}
\label{eq:crit-1}
\frac{\delta J}{\delta \varepsilon(t)}
= \left\langle \nabla J(U_T), \frac{\delta U_T}{\delta \varepsilon(t)} \right\rangle = 0 ,
 \ \ \ \forall t \in [0,T] ,
\end{equation}
where $\nabla J(U_T)$ is the gradient of $J$ at $U_T$, the Jacobian matrix $\delta U_T / \delta \varepsilon(t)$ is the first-order functional derivative of $U_T$ with respect to the control field, and $\langle \cdot , \cdot \rangle$ is the Hilbert-Schmidt inner product. A critical point of $J$ is \emph{regular} if the Jacobian $\delta U_T / \delta \varepsilon(t)$ is of full rank, and \emph{singular} if $\delta U_T / \delta \varepsilon(t)$ is rank-deficient. If conditions (1) and (3) for a landscape free of local optima are satisfied, i.e, the system is controllable and the control field is unconstrained, then none of the regular landscape critical points are local optima \cite{RabitzHsiehRosenthal2005PRA, HsiehRabitz2008PRA, HoDominyRabitz2009PRA, HsiehWuRabitzLidar2010, RabitzHoHsieh2006PRA, RabitzHsiehRosenthal2006JCP, WuRabitzHsieh2008JPA, HsiehWuRabitz2009JCP, ChakrabartiRabitz2007, Brif2010NJP, Brif2012ACP}. No such result has been demonstrated for singular critical points, nor, at present, is there an analytical method to determine whether there are singular critical points on the landscape corresponding to a particular control problem. However, a recent numerical study \cite{WuLongDominy2012} described an algorithm capable of locating singular critical points; various control problems were studied and none of the detected singular points trapped gradient searches. This result indicates that the overwhelming majority of singular critical points are not local optima. Another pair of recent works \cite{PechenTannor2011, *PechenTannor2011Comm, *PechenTannor2011CommResp, FouquieresSchirmer2013} showed that, for several specially constructed combinations of control objective and Hamiltonian, a singular critical point at $\varepsilon(t) = 0$ is a second-order trap.  For a maximization problem, a critical point is a second-order trap if the Hessian matrix of the second functional derivatives of $J$ with respect to the field,
\begin{equation}
\label{eq:hess}
\mathsf{H}(t,t') = \frac{\delta^2 J}{\delta \varepsilon(t) \delta \varepsilon(t')} ,
\end{equation}
is negative semidefinite.  Such a trap is not necessarily a local maximum of the landscape, since higher-order functional derivatives may be indefinite \cite{PechenTannor2012IJC}, but it can in principle prevent a simple gradient search from finding a globally maximal solution. However, a subsequent computational study \cite{RivielloBrif2014} examined the same control problems as \cite{PechenTannor2011, PechenTannor2011Comm, PechenTannor2011CommResp, FouquieresSchirmer2013} and found that the second-order traps only attract search trajectories that originate very close to them (i.e., at fields which are several orders of magnitude weaker than the optimal ones) and thus are very unlikely to affect gradient-based optimizations under realistic searching conditions. In this work, we nonetheless assume, for the sake of simplicity, that condition (2) is satisfied and that there are no singular critical points on the control landscape.

When a control problem satisfies conditions (1) and (2), the corresponding landscape is free of fundamental traps. However, constraints on the control field violate condition (3) and can interfere with optimization. Unlike the first two conditions, some constraints are unavoidable; for example, in OCEs with lasers, the number of available control variables is determined by the design of the pulse shaper and bandwidth limitations are dictated by the optical source. These restrictions were discussed in early experimental studies \cite{PearsonWhite2001}. OCT simulations generally discretize the system evolution, which also constrains the control fields that can be generated. In this paper, we focus on the subset of \emph{severe} constraints, i.e., those that prevent achievement of the target objective by introducing local optima onto the control landscape.  It has been shown, however, that even more mild constraints can have a significant effect on OCT optimizations, e.g., by increasing the search effort \cite{GollubKowalewski2008, PalaoKosloff2008, PalaoReich2013, LapertTehini2009, LapertSalomon2012, MooreBrif2012}.

Several approaches have been taken to address the presence of control constraints. Special algorithms that facilitate successful optimization when the control field has significant spectral constraints have been introduced, for problems such as population transfer in a one-dimensional asymmetric double well \cite{WerschnikGross2005} and molecular alignment \cite{LapertTehini2009}. Other works have explored the effect of a specific constraint on OCT optimization; time-optimal control, the problem of achieving a target objective in the minimum possible time, has received the greatest attention \cite{Khaneja2001, Khaneja2002, KhanejaHeitmann2007, Masanes2002, BoscainChitour2005, SchulteSporl2005, NielsenDowling2006, Carlini2007, Carlini2011, KoikeOkudaira2010, CanevaCalarco2011, MooreBrif2012}, and constraints on the number of field components have also been investigated \cite{MooreRabitz2012}. In this work, we perform extensive OCT simulations to evaluate constraints whose effects on the success of gradient optimization have not previously been examined, identifying values of each constrained parameter beyond which some or all of a set of searches fail to optimize. We also expand upon these prior studies to include new systems and objectives.

\section{Methodology}
\label{sec:methods}

\subsection{Quantum control objectives and corresponding landscape topology}
\label{sec:structure}

The aim of OCT and OCEs is to find a control field $\varepsilon(t)$ that corresponds to the global maximum (or minimum) of an objective functional $J[\varepsilon(t)]$. The OCT simulations in this paper target three common quantum control goals:
\begin{enumerate}[(I)]
\item The \emph{state-transition objective} is to maximize the probability of a transition between initial and final pure states $| i \rangle$ and $| f \rangle$ at time $T$:
\begin{equation}
\label{eq:pif}
J_P = | \langle f | U_T | i \rangle | ^2 .
\end{equation}
\item The \emph{observable objective} is to maximize the expectation value of a quantum observable $\theta$ at time $T$:
\begin{equation}
\label{eq:tro}
J_{\theta} = \mathrm{Tr} \left( U_T^\dagger \theta U_T \rho_0 \right) .
\end{equation}
\item The \emph{evolution-operator objective} is to minimize the distance between $U_T$, the unitary evolution operator at time $T$, and a target unitary transformation $W$:
\begin{equation}
\label{eq:w}
J_W = \frac{1}{2} - \frac{1}{2N} \Re\, \mathrm{Tr} \left( W^{\dagger} U_T \right) .
\end{equation}
\end{enumerate}The state-transition objective $J_P$ is a special case of the observable objective $J_{\theta}$ for which $\rho_0 = | i \rangle \langle i |$ and $\theta = | f \rangle \langle f |$, i.e., $\rho_0$ and $\theta$ are projectors onto the states $| i \rangle$ and $| f \rangle$, respectively. Unless noted otherwise, the simulations in this work only consider $\rho_0$ and $\theta$ that are diagonal in the eigenbasis of $H_0$, an assumption that still permits a fully general analysis of the control landscape topology \cite{WuRabitzHsieh2008JPA}.

The landscape analysis for objectives (I) -- (III) can be performed in either the \textit{dynamic formulation}, in which the control landscape $J = J[\varepsilon(t)]$ is defined on the $L^2$ space of control fields, or the \textit{kinematic formulation}, in which the control landscape $J = J(U_T)$ is defined on the unitary group U$(N)$. If the Jacobian $\delta U_T/\delta \varepsilon(t)$ is of full rank at a critical point $\delta J / \delta \varepsilon(t) = 0$ in the dynamic formulation, then the final-time propagator $U_T$ corresponding to that control field must also satisfy the kinematic critical point condition $\nabla J(U_T) = 0$. In general, there exist many critical control fields $\varepsilon(t)$ that correspond to the same critical propagator $U_T$. Additionally, at a regular critical point, the number of positive and negative eigenvalues in the Hessian spectrum are the same in the dynamic and kinematic formulations \cite{WuRabitzHsieh2008JPA}. Therefore, if conditions (1) -- (3) for a trap-free landscape are met, then the kinematic and dynamic formulations of the control landscape have the same topology.

Under the assumption that conditions (1) -- (3) are satisfied, the analysis of the landscape topology for control objectives (I) -- (III) has been performed in the kinematic formulation \cite{Brif2012ACP} and all critical points have been characterized. The landscape $J_P(U_T)$ for pure-state transition control has two critical points that correspond to the global maximum at $J_P = 1$ and the global minimum at $J_P = 0$, respectively \cite{RabitzHsiehRosenthal2004, RabitzHoHsieh2006PRA}. In general, the landscape $J_{\theta}(U_T)$ for observable control has a global maximum and a global minimum as well as other critical points that are shown to be saddles by the analysis of the Hessian spectrum \cite{HoRabitz2006JPPA, WuRabitzHsieh2008JPA}. The values of the objective $J_{\theta}$ that correspond to critical points are determined by the eigenvalues of the initial density matrix $\rho_0$ and the target observable $\theta$. When $\rho_0$ and $\theta$ are both pure-state projectors, the observable landscape has the same topology as the state-transition landscape, with no saddle points.  When $\rho_0$ and $\theta$ are of full rank, the observable landscape contains $N!$ critical points, of which $N! - 2$ are saddles \cite{HoRabitz2006JPPA, WuRabitzHsieh2008JPA}. For $\rho_0$ and $\theta$ with other eigenvalue spectra, the observable landscape has fewer than $N! - 2$ saddles. For evolution-operator control, the landscape $J_W(U_T)$ has $N+1$ critical points corresponding to the objective values $J_W = 0, 1/N, 2/N, ... ,1$. The global minimum and maximum correspond to the objective values $J_W = 0$ and $J_W = 1$, respectively, while the other critical points are saddles \cite{HsiehRabitz2008PRA, HoDominyRabitz2009PRA}. We will denote the objective values corresponding to the global maximum and minimum of a control landscape as $J_{\max}$ and $J_{\min}$, respectively.

\subsection{The optimization procedure}
\label{sec:contproc}

OCT simulations and OCEs have used a variety of optimization algorithms to find globally optimal controls for the objectives defined in Eqs.~\eqref{eq:pif} -- \eqref{eq:w} \cite{Brif2010NJP, Brif2012ACP}. \emph{Global methods}, such as genetic algorithms, sample a large region of the control space stochastically and can therefore avoid trapping at local optima at the expense of a lower efficiency. \emph{Local methods} include the gradient-based and simplex algorithms, the former of which have been employed with great success in OCT simulations due to the absence of local traps when conditions (1) -- (3) are satisfied. Gradient-based methods also have several properties that render them ideal for identifying the local optima that may arise when one of the three conditions is violated. They are ``myopic'', i.e, any step taken by the algorithm is dictated by the geometry of the control landscape at the current control field, and they are deterministic, i.e., the algorithm will always take the same step at the same point on a given landscape. Therefore, we employ a gradient-based algorithm in the OCT simulations in this work. Atomic units are used throughout this paper. 

Each numerical optimization in this paper is parameterized in terms of a dimensionless index $s$, which denotes the changes made to the control field as the search proceeds. Therefore, we write the control field as $\varepsilon(s,t)$, where the value $s = 0$ corresponds to the initial field $\varepsilon_0(t)$. Successive control fields ($s > 0$) are found by solving the initial value problem
\begin{equation}
\label{eq:searchalg}
\frac{\partial \varepsilon(s,t)}{\partial s} = \gamma \frac{\delta J[\varepsilon(s,t)]}{\delta \varepsilon(s,t)} , 
\quad \varepsilon(0,t) = \varepsilon_0(t),
\end{equation}
where $\gamma$ is a positive (negative) constant when maximizing (minimizing) $J$. Using the following result \cite{HoRabitz2006JPPA}:
\begin{equation}
\label{eq:UT-deriv}
\frac{\delta U_T}{\delta \varepsilon(t)} = \frac{i}{\hbar} U_T \mu(t) , \quad \mu(t) = U^{\dag}(t) \mu U(t) ,
\end{equation}
one can apply the chain rule, as in Eq.~\eqref{eq:crit-1}, to calculate the functional derivative $\delta J / \delta \varepsilon(s,t)$ in Eq.~\eqref{eq:searchalg} for the quantum control objectives in Eqs.~\eqref{eq:pif} -- \eqref{eq:w}. The result is \cite{HoRabitz2006JPPA, RabitzHoHsieh2006PRA, HoDominyRabitz2009PRA, MooreRabitz2012}:
\begin{subequations}
\label{eq:grad}
\begin{align}
& \frac{\delta J_P}{\delta \varepsilon(t)} 
= \frac{2}{\hbar} \Im\, \left[ \langle f | U_T | i \rangle \langle i | \mu(t) U_T^{\dag} | f \rangle \right] , \\
& \frac{\delta J_{\theta}}{\delta \varepsilon(t)} 
= \frac{2}{\hbar} \Im\, \mathrm{Tr} \left[ U_T^\dagger \theta U_T \rho_0 \mu(t) \right] , \\
& \frac{\delta J_W}{\delta \varepsilon(t)} 
= \frac{1}{2N \hbar} \Im\, \mathrm{Tr} \left[ W^{\dagger} U_T \mu(t) \right] .
\end{align}
\end{subequations}
Having calculated the functional derivative of $J$ with respect to the control field, we numerically solve Eq.~\eqref{eq:searchalg} using the MATLAB routine \texttt{ode45} \cite{matlab}, which implements a variable-step-size fourth-order Runge-Kutta gradient method. Searches using \texttt{ode45} must specify the \emph{absolute error tolerance} $\tau$, a positive quantity that influences the determination of the step size at each algorithmic iteration.  The simulations in this paper use the value $\tau = 10^{-8}$ unless otherwise stated; a prior numerical study \cite{RivielloBrif2014} indicates that this choice of $\tau$ generally leads to excellent solutions of Eq.~\eqref{eq:searchalg}. The optimization is considered to have converged successfully when the search reaches a control field $\varepsilon(s_f,t)$ corresponding to an objective value $J \geq (J_{\max} - \eta)$ (for maximization of $J$) or $J \leq (J_{\min} + \eta)$ (for minimization of $J$). Smaller values of the convergence parameter $\eta$ demand greater accuracy from the optimal control field. In this paper, we use the value $\eta = 0.001 \cdot (J_{\max} - J_{\min})$. The \emph{search effort} is defined as the number of iterations required for the optimization to converge and is an important indicator of algorithmic efficiency.

Controls that satisfy the critical point condition in Eq.~\eqref{eq:crit} are in principle continuous. However, numerical optimizations typically represent $\varepsilon(t)$ as a piecewise-constant function; in this work, the control field is defined over $L$ equal intervals of time, each of length $\Delta t = T/L$:
\begin{equation}
\label{eq:discrete}
\varepsilon(t) = \{ \varepsilon_l | t \in (t_{l-1},t_l] \}_{l=1}^L,
\end{equation}   
where $t_l = l \Delta t$. With the control field defined in this way, Eq.~\eqref{eq:schro} can be numerically integrated by calculating a series of incremental evolution operators, each of which propagates the system over one of the $L$ constant-field intervals:
\begin{equation}
\label{eq:incrementU}
U(t_l,t_{l-1}) = \exp \left[ -\frac{i}{\hbar} (H_0 - \mu \varepsilon_l) \Delta t \right] ,
\end{equation}
and constructing the evolution operator $U(t_l,0)$ as a product of these incremental propagators:
\begin{equation}
\label{eq:calcu}
U(t_l,0) = U(t_l,t_{l-1}) \cdots U(t_2,t_1) U(t_1,0) , \\
\end{equation}
where the final-time evolution operator is $U_T = U(t_L,0)$. The control field discretization $\Delta t$ must be sufficiently small in order for this piecewise-constant field to accurately approximate a continuous one.  A large value of $\Delta t$ may severely constrain the control field, as discussed in Sec.~\ref{sec:deltat}.

The specific optimization procedure depends on the choice of control variables. There are many possible choices, but this work uses two common ones:
\begin{enumerate}[(i)]
\item The control variables are the $L$ field values $\{\varepsilon_l\}$ defined in Eq.~\eqref{eq:discrete}. They are real-valued and independently addressable. The simulations in this work begin with a vector of initial field values $\{\varepsilon_l(0)\}$:
\begin{subequations}
\label{eq:field-init-1}
\begin{align}
& \varepsilon_l(0) = A(t_l) \sum_{m=1}^M a_m \cos (\omega_m t_l) , \label{eq:field-init-1a} \\
& A(t_l) = A_0 \exp \left[ -(t_l-T/2)^2 / (2 \zeta^2) \right] , \label{eq:field-init-1b}
\end{align}
\end{subequations}
where $A(t_l)$ is the Gaussian envelope function whose width is determined by the positive parameter $\zeta$. We use $\zeta = T / 10$, which enforces the conditions that $\varepsilon_0(t) \approx 0$ at $t = 0$ and $t = T$, and $M = 20$ except when otherwise noted. The frequencies $\{ \omega_m \}$ are randomly selected from a uniform distribution on $[\omega_{\min},\omega_{\max}]$, where $\omega_{\min}$ and $\omega_{\max}$ are the smallest and largest transition frequencies in $H_0$, respectively. The amplitudes $\{ a_m \}$ are randomly selected from a uniform distribution on $[0, 1]$. The normalization constant $A_0$ is chosen so that the fluence, 
\begin{equation}
\label{eq:fluence}
F = \| \varepsilon \|_2^2 = \int_0^T \varepsilon^2(t) d t ,
\end{equation}
of each initial field has the same value $F_0$. After the initialization (i.e., for $s > 0$), the field values $\{ \varepsilon_l(s) \}$ are allowed to vary independently at each step of the optimization algorithm, and the optimization proceeds by solving a discrete analog of Eq.~\eqref{eq:searchalg}:
\begin{equation}
\label{eq:searchalg-discr-1}
\frac{\partial \varepsilon_l(s)}{\partial s} = \gamma \frac{\delta J}{\delta \varepsilon_l(s)} = \gamma \Delta t \frac{\partial J}{\partial \varepsilon(t_l)}. 
\end{equation}
This flexible set of control variables allows the field fluence to vary freely during the search.
\item The control variables are the phases $\{ \phi_m \}$ of $M$ spectral components of the field, which has the form 
\begin{equation}
\label{eq:field-1}
\varepsilon(t) = A(t) \sum_{m=1}^M \cos (\omega_m t + \phi_m) . 
\end{equation}
The envelope function $A(t)$ and the frequencies $\{ \omega_m \}$ are chosen at the beginning of the search and remain fixed throughout the optimization; additionally, the amplitude of the $m$-th term in Eq.~\eqref{eq:field-1} remains at 1.0. Thus, this form is constrained even when $M$ is large. The field is still discretized into $L$ intervals as in Eq.~\eqref{eq:discrete}. $A(t)$ is defined as in Eq.~\eqref{eq:field-init-1b}. The gradient-based algorithm generates an evolving phase vector $\{\phi_m (s)\}$ along the search trajectory by solving the equation \cite{MooreRabitz2012}:
\begin{equation}
\label{eq:searchalg-discr-2}
\frac{\partial \phi_m(s)}{\partial s} = \gamma \frac{\partial J}{\partial \phi_m(s)} , 
\end{equation}
where elements of the gradient vector are obtained from
\begin{equation}
\frac{\partial J}{\partial \phi_m} = \int_{0}^T \frac{\delta J}{\delta \varepsilon(t)} 
\frac{\partial \varepsilon(t)}{\partial \phi_m} d t ,
\end{equation}
and the search starts from a vector of initial phase values, $\{\phi_m(0)\}$, each of which are randomly chosen from the interval $[0,2 \pi]$. Since the envelope function and the amplitudes of the field components are fixed, the fluence remains very close to its initial value $F_0$ throughout the optimization.
\end{enumerate}
Each choice of control variables, including others beyond those above, has its own advantages and limitations. Choice (i) makes it possible to represent arbitrary shapes of the control pulse as $L$ increases. Choice (ii) is more representative of a pulse shaper's output, but its form is inherently constrained as the amplitude of each field component is fixed.

\section{Effects of severe control field constraints}
\label{sec:resources}

Several OCT studies have shown that violating condition (3) by limiting the number of control variables \cite{MooreRabitz2012} or the control period $T$ \cite{Khaneja2001, Khaneja2002, KhanejaHeitmann2007, Masanes2002, BoscainChitour2005, SchulteSporl2005, NielsenDowling2006, Carlini2007, Carlini2011, KoikeOkudaira2010, CanevaCalarco2011, MooreBrif2012} can prevent the achievement of a globally optimal solution. In this section, we investigate the practical effects of imposing various types of constraints. It is not possible to avoid constraints altogether, as they result from any of the limitations on experimental or computational parameters that are invariably present in OCE and OCT. Constraints do not necessarily interfere with optimization \cite{PalaoReich2013}, but severe constraints introduce artificial local optima and saddles to the control landscape. Thousands of successful simulations in the quantum control literature were facilitated by having only relatively mild constraints on the control field. This section cannot serve as an exhaustive rubric for evaluating whether a specific control scheme is amenable to successful optimization, nor is it a comprehensive list of significant control constraints. However, the simulations below examine several common constraints that, when sufficiently severe, are very likely to impede a gradient search. OCEs and OCT simulations almost always involve multiple constraints, which may have a cumulative effect on the success of an optimal search.  In this section, we study each constraint as independently as possible by introducing only one severe constraint for each set of simulations. 

\subsection{Representation of the control field and system dynamics}
\label{sec:deltat}

The numerical representation of the control field as a piecewise-constant function of time, as in Eq.~\eqref{eq:discrete}, and the corresponding discretized unitary system evolution in Eq.~\eqref{eq:calcu}, is a common practical procedure in OCT. This discrete representation of $\varepsilon(t)$ constrains the theoretically continuous field. Gradient searches have been observed to fail with a large time discretization interval $\Delta t$, and subsequently optimize when $\Delta t$ was reduced \cite{MooreRabitz2011}. These results suggest that a sufficiently small $\Delta t$ is essential for successful optimization. In particular, $\Delta t$ must be small enough to resolve all the features required of an optimal field. While one cannot \emph{a priori} predict the pulse shapes required to optimize a particular objective, high-frequency transitions essential to the optimal field can only be resolved using a finer time discretization. Many laboratory experiments use pulse shapers with analogously discretized elements $\Delta \omega$ in the frequency domain.

We performed numerical optimizations of $J_\theta$, using a range of $\Delta t$ values, on the quantum system 
\begin{subequations}
\label{eq:genhamdip}
\begin{align}
& H_0 = \sum_{j=0}^{N-1} \lambda j(j+1) | j \rangle \langle j |,  \\
& \mu = \sum_{j \neq k}^{N-1} \frac{D^{|j-k|}}{D} |j \rangle \langle k | ,
\end{align}
\end{subequations}
with $N = 6$, $\lambda = 1$ and $D = 0.5$. In these optimizations, we used choice (i) of control variables described in Sec.~\ref{sec:contproc}, i.e., the  $L$ field values $\{\varepsilon_l(s)\}$. The optimization goal was to maximize the objective $J_\theta$, with the initial state and target observable selected as $\rho_0 = \sum_{j=0}^1 p_j | j \rangle \langle j |$ with $p_0 = 0.6$ and $p_1 = 0.4$, and $\theta = \sum_{k=3}^5 \theta_k | k \rangle \langle k |$ with $\theta_3 = 0.1$, $\theta_4 = 0.2$, and $\theta_5 = 0.7$. The final time was $T = 50$ and $\Delta t = 50/L$, where $L$ is the number of intervals into which the time period $[0,T]$ is divided. 1000 optimization runs were performed for each value of $\Delta t$ over the range $0.098 \leq \Delta t \leq 0.625$ ($511 \geq L \geq 79$).

\begin{figure}[htbp]
\centering
\includegraphics[width=8cm]{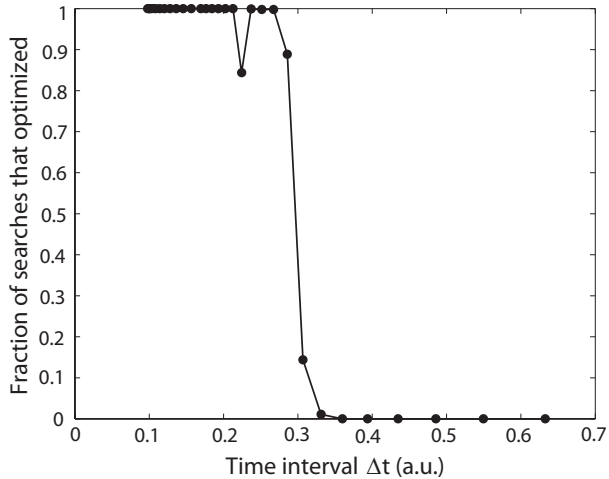}
\caption{The fraction of searches that optimized successfully, as a function of the time discretization interval $\Delta t$, using control form (i). 1000 optimization runs were performed for each $\Delta t$ value.}
\label{fig:tpts}
\end{figure}
Figure~\ref{fig:tpts} shows that the fraction of searches that optimized clearly depends on $\Delta t$. For $\Delta t > 0.213$, at least one search failed to optimize, while for $\Delta t > 0.331$, all searches failed to optimize. These results indicate that constraint-induced traps begin to emerge on the control landscape for $\Delta t > 0.213$, while for $\Delta t > 0.331$ the global optimum may be unreachable. These two values help to quantify the effect of constraining the time discretization interval. Thus, the choice of a large $\Delta t$ severely constrains the control field, whereas gradient searches will generally optimize when $\Delta t$ is sufficiently small.

\subsection{Number of control variables}
\label{sec:numcontrols}

\begin{figure}[htbp]
\centering
\includegraphics[width=8cm]{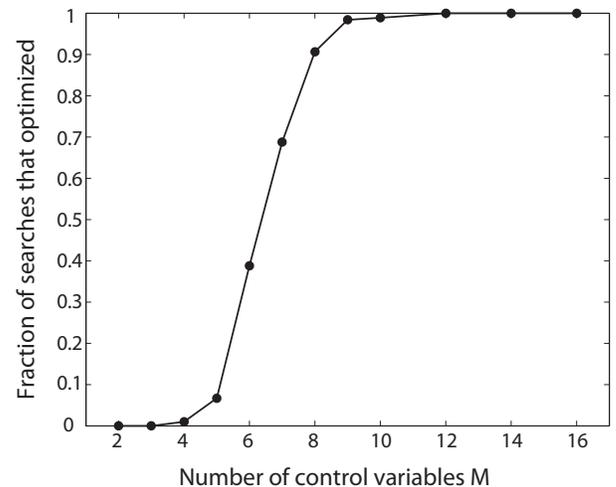}
\caption{The fraction of searches that optimized successfully, as a function of the number of control variables $M$, using the control form (ii) in Eq.~\eqref{eq:field-1}. The initial field fluence  was $F_0 = 10^3$, and} 1000 optimization runs were performed for each value of $M$.
\label{fig:numcontrols}
\end{figure}

Control landscape analysis shows that for the state-transition objective the gradient $\delta J_P / \delta \varepsilon(t)$ can be constructed from at most $2N-2$ independent basis functions \cite{HsiehHoRabitz2008CP} (the so-called natural basis), and that the Hessian matrix $\mathsf{H}(t,t')$ at a globally optimal solution contains no more than $2N-2$ negative eigenvalues \cite{MooreRabitz2011, ShenHsiehRabitz2006JCP}.  It was shown \cite{HsiehHoRabitz2008CP} that simulations of the objective $J_P$ using the natural basis will optimize successfully with a gradient-based method similar to the one described in Sec.~\ref{sec:contproc}.  In this section, we chose a different set of control variables and performed optimizations of $J_P$ to investigate the degree to which constraining the number of control variables prevents gradient-based searches from optimizing. We used the control form (ii) in Sec.~\ref{sec:contproc}, so the control variables were the phases $\{ \phi_m \}$, whereas the frequencies $\{ \omega_m \}$, the amplitudes $\{ a_m \}$, and the envelope function $A(t)$ were fixed. The frequencies were set to integer values $\omega_m = m$ and the amplitudes were identically $a_m = 1$ $\forall m$. The control period $T = 50$ was divided into $L = 1023$ intervals, and the initial field fluence was $F_0 = 10^3$. The simulations were performed on the system from Eqs.~\eqref{eq:genhamdip}, with $N = 4$, $\lambda = 1$, and $D = 0.9$. The goal was to maximize $J_P$ for the transition $| 0 \rangle \to | 3 \rangle$. 1000 optimization runs were performed for each value of $M$ (the number of control variables) over the range $3 \leq M \leq 16$. Statistical results from these simulations are summarized in Fig.~\ref{fig:numcontrols}, which illustrates that at least one search failed for $M < 12$ and all searches failed for $M < 4$. These data confirm that an insufficient number of control variables (here, in the spectral domain) is a severe constraint.

Unlike the natural basis described in \cite{HsiehHoRabitz2008CP}, choice (ii) does not ensure successful optimization when $M = 2N -2$  control variables are used; similar behavior was observed in an earlier work \cite{MooreRabitz2012} with a different system. This result confirms that different choices of control parameterization may require a distinct number of variables in order to optimize successfully. In addition, choice (ii) of the control variables contains parameters that themselves must be chosen carefully in order for optimizations to be successful. The simulations in Fig.~\ref{fig:numcontrols} included field components $\{ \omega_m \}$ resonant with transitions in $H_0$, and they optimized when $M$ was sufficiently large.  It has been shown \cite{MooreRabitz2012} that simulations using control fields with no resonant field components are much more likely to fail than those using fields with resonant components. However, the intuitively appealing choice of including \emph{only} resonant field components does not necessarily improve optimization success. In Fig.~\ref{fig:numcontrols}, 67 of 1000 simulations with $M = 5$ optimized successfully; these runs used a combination of resonant and non-resonant field frequencies, with $\omega_m = m$. We also performed 1000 simulations on the same control problem, but instead used five field components corresponding to the resonant transitions in $H_0$. None optimized successfully.  Therefore, while it is clear that an improper choice of variables can severely constrain the control field, there is no known method \textit{a priori} to be certain that a set of variables is inappropriate.

\subsection{Duration of control pulse}
\label{sec:T}

Theoretical analysis and numerical simulations have both shown that a sufficiently large control time $T$ is necessary in order to generate an optimal control; for example, a recent computational study evaluated the minimum time required to optimize the objective $J_W$ \cite{MooreBrif2012} for two-, three-, and four-qubit coupled-spin model systems. In particular, the CNOT, SWAP, and quantum Fourier transform (QFT) gates were chosen as the target unitary transformations. For each control problem, it was shown that some minimum control time is necessary for successful optimization.

Another control problem \cite{FouquieresSchirmer2010} utilizes six control fields with the objective of minimizing $\tilde{J}_W = 1 - \frac{1}{N} \left| \mathrm{Tr} \left( W^{\dag} U \right) \right|$, a phase-independent form \cite{MooreBrif2012} of the evolution-operator objective $J_W$ (see Eq.~\eqref{eq:w}), for an eight-level system consisting of three Ising-coupled qubits:
\begin{equation}
\label{eq:qftham}
\begin{split}
&2 H(t) = Z_1 Z_2 + Z_2 Z_3 + \varepsilon_1(t) X_1 + \varepsilon_2(t) Y_1 \\
&+ \varepsilon_3(t) X_2 + \varepsilon_4(t) Y_2 + \varepsilon_5(t) X_3 + \varepsilon_6(t) Y_3 ,
\end{split}
\end{equation}
where the operators $X_1 = \sigma_x \otimes I \otimes I$, $Y_2 = I \otimes \sigma_y \otimes I$, $Z_3 = I \otimes I \otimes \sigma_z$, etc. The target unitary transformation is the three-qubit QFT gate: 
\begin{equation}
\label{eq:qftW}
W = \sum_{j,k=1}^8 \frac{\exp(2 \pi i (n + \frac{1}{4})/8)}{\sqrt{8}} \xi^{jk} | j \rangle \langle k | ,
\end{equation}
where $\xi = \exp (-2 \pi i/8)$ and $n$ is an integer. In \cite{FouquieresSchirmer2010}, 1000 OCT optimizations of this problem, with $n=5$, were performed using a control period $T = 8$ divided into $L = 140$ intervals. A small fraction of them became trapped at suboptimal fidelities. In another work \cite{RivielloBrif2014}, optimizations of $\tilde{J}_W$ on the same system were performed for $T = 6,7,8,9,10$; all runs failed to optimize for $T = 6$ and all runs succeeded for $T = 10$, leading to the conclusion that the smaller choices of $T$ severely constrain the control field.

In this work, we performed optimizations of the objective $J_W$ with different control systems and unitary targets than in these prior works. We used the control variables (i) described in Sec.~\ref{sec:contproc}, and the Hamiltonian was defined as in Eq.~\eqref{eq:genhamdip}, with $N = 5$, $\lambda = 1$, and $D = 0.9$. To ensure controllability, i.e., that any $W \in U(N)$ can be generated by the Hamiltonian evolution, it is required that $\mathrm{Tr} (\mu) \neq 0$ \cite{DAlessandro2007}. In order to satisfy this condition, the diagonal dipole elements were set as $\left \langle j | \mu | j \right \rangle = 1$ $\forall j$ in these simulations. Quasirandom target unitary transformations $W_j$ were chosen by first constructing Hermitian matrices $A_j$; the real and imaginary part of each element of $A_j$ was randomly generated on the interval $[0, 2 \pi]$, subject to the restrictions of hermiticity.  The targets $W_j$ were then generated using the relation
\begin{equation}
\label{eq:expiA}  
W_j = \exp \left ( i A_j \right ) .
\end{equation}
The optimizations in this section were performed on two target transformations, 
\begin{widetext}
\begin{subequations}
\label{eq:w1w2}
\begin{align}
W_1 &= \begin{pmatrix} 0.456 - 0.034i & 0.064 + 0.711i & 0.055 + 0.108i & 0.222 + 0.163i & 0.409 - 0.154i \\ -0.031 - 0.246i & -0.418 + 0.232i &  0.621 - 0.342i & -0.137 - 0.417i & -0.101 - 0.067i \\ -0.399 - 0.008i & -0.236 + 0.003i & -0.076 + 0.216i & -0.327 + 0.139i &  0.278 - 0.727i \\  -0.485 - 0.329i &  0.112 + 0.326i &  0.198 + 0.521i & -0.123 + 0.156i & -0.071 + 0.427i  \\  0.075 + 0.471i &  0.225 + 0.190i & -0.074 + 0.337i & -0.347 - 0.668i & -0.035 + 0.004i  \end{pmatrix} \\
W_2 &= \begin{pmatrix}  0.131 + 0.215i & -0.005 - 0.039i & -0.121 + 0.034i &  0.292 + 0.332i  & 0.603 + 0.601i  \\  0.084 - 0.732i &  0.317 - 0.420i &  0.122 + 0.240i & -0.232 + 0.059i &  0.123 + 0.187i \\  0.119 - 0.023i  & 0.082 + 0.098i  & 0.749 - 0.486i  & 0.087 - 0.318i &  0.055 + 0.245i  \\  0.083 - 0.584i & -0.239 + 0.423i & -0.126 - 0.214i &  0.474 + 0.287i & -0.217 + 0.031i  \\  -0.105 - 0.143i & -0.241 + 0.641i &  0.099 + 0.208i & -0.575 - 0.017i &  0.330 + 0.080i  \end{pmatrix} ,
\end{align}
\end{subequations}
\end{widetext}
that were chosen in this way. The control period $T$ was divided into $L = 128$ intervals, and 100 simulations were performed for each target and for each value of $T$ over the range $1 \leq T \leq 4$.    
\begin{figure}[htbp]
\centering
\includegraphics[width=8cm]{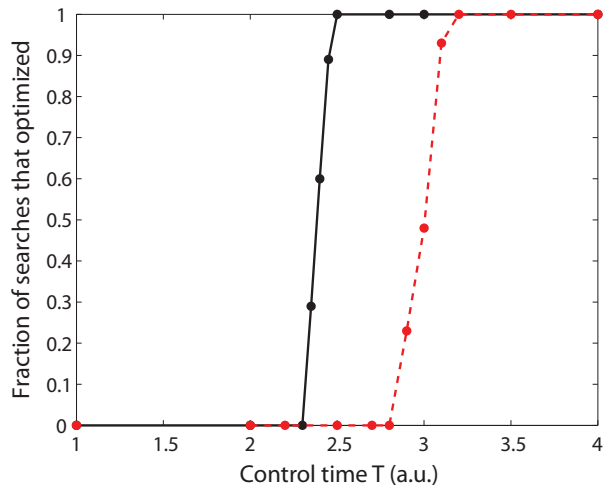}
\caption{(Color online) The fraction of searches that optimized successfully as a function of the control pulse duration $T$, for target unitary transformations $W_1$ (solid black line) and $W_2$ (dashed red line). 100 optimization runs were performed for each value of $T$.}
\label{fig:T}
\end{figure}
Figure~\ref{fig:T} shows that for the target $W_1$, at least one search failed for $T \leq 2.45$ and all searches failed for $T \leq 2.3$; for the target $W_2$,  at least one search failed for $T \leq 3.1$ and all searches failed for $T < 2.8$. These results indicate that insufficient $T$ is a severe control constraint in optimizations not only for the previously-studied multi-qubit systems \cite{MooreBrif2012, RivielloBrif2014}, but also for multilevel systems as defined in Eq.~\eqref{eq:genhamdip}. Threshold values of $T$ can be identified in these latter systems, and most importantly, distinct threshold values for $T$ exist for each target unitary transformation.  The choice of $T = 2.8$, for example, resulted in the success of all optimizations targeting the transformation $W_1$, but the failure of all optimizations targeting $W_2$.  This point emphasizes that the distinction between a severe and mild constraint is highly problem-dependent and can be established by a single parameter in otherwise similar optimizations.   

\subsection{Strength of the control field}
\label{sec:fieldstrength}

A control field of insufficient strength can impede the achievement of the control objective.  In this work, we use the field fluence $F$ [see Eq.~\eqref{eq:fluence}] as a measure of the strength of the control field. It is often desirable that control simulations and experiments achieve an optimal field while also minimizing the fluence. This is commonly attempted by adding a fluence penalty term to the objective functional, which is a constraint discussed in Sec.~\ref{sec:compos}.  However, searches that use choice (ii) of the control variables (see Sec.~\ref{sec:contproc}) also constrain the field since the fluence cannot increase significantly during the search. To investigate the effect of fluence constraints on the results of an optimization, we performed two sets of optimizations of $J_P$ using the same control system but different choices of variables. In both sets of runs, the system from Eq.~\eqref{eq:genhamdip} ($N = 4$, $\lambda = 1$, $D = 0.9$) was used, the transition $| 0 \rangle \to | 3\rangle$ was targeted, and the control period $T = 50$ was divided into $L = 1023$ intervals.

\begin{figure}[htbp]
\centering
\includegraphics[width=8cm]{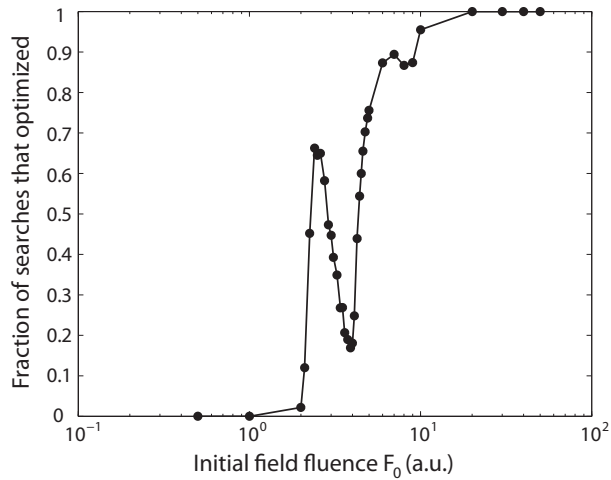}
\caption{The fraction of searches that optimized successfully, as a function of the initial fluence $F_0$, for simulations that used choice (ii) of the control variables. 1000 optimization runs were performed for each $F_0$ value, for $0.5 \leq F_0 \leq 50$. The field fluence remains very close to $F_0$ during the search.}  
\label{fig:fieldstrength}
\end{figure}

The first set of simulations used choice (ii) of the controls, with $M = 16$ phase component variables. The frequencies were set to integer values $\omega_m =  m$, and the initial field fluence $F_0$ had a pre-selected value ranging over $0.5 \leq F_0 \leq 50$. As a result of the choice of variables, the fluence $F$ remains very close to $F_0$ for the entire optimization. 1000 optimizations were performed for each value of $F_0$. Figure~\ref{fig:fieldstrength} shows a clear relationship between the initial fluence $F_0$ and the fraction of searches that failed to optimize; at least one search failed for $F_0 < 30$ and all searches failed for $F_0 < 2$.  The non-monotonic behavior in Fig.~\ref{fig:fieldstrength} is probably an artifact of the particular control parameterization and may also be related to an oscillatory dependence of the optimal-field fluence on the control duration, which has been previously observed in \cite{MooreBrif2012, BrifGrace2013Talk}.

\begin{figure}[htbp]
\centering
\includegraphics[width=8cm]{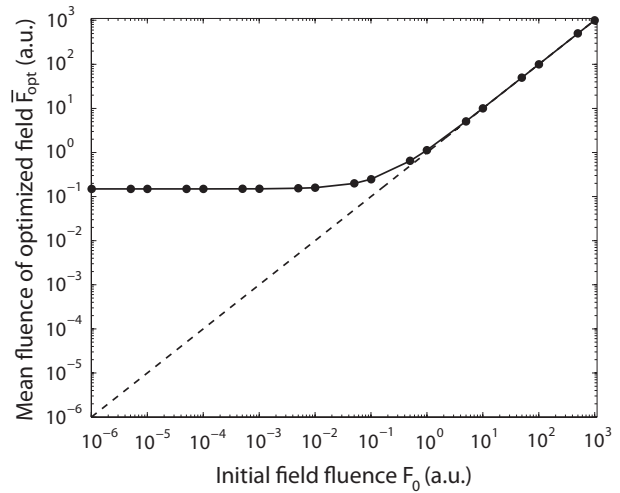}
\caption{The mean fluence $\overline{F_{\mathrm{opt}}}$ of the optimized field as a function of the initial fluence $F_0$, for simulations that used choice (i) of the control variables. Twenty optimizations were performed for each $F_0$ value. The dashed line indicates where $\overline{F_{\mathrm{opt}}} = F_0$.}  
\label{fig:initfinfluence}
\end{figure}

The second set of simulations used choice (i) of the control variables. One hundred runs were performed for each $F_0$ value over the range $10^{-6} \leq F_0 \leq 10^3$, and every search succeeded. For each $F_0$ value, the mean fluence $\overline{F_{\mathrm{opt}}}$ of the twenty optimized fields was computed. These statistical results are summarized in Fig.~\ref{fig:initfinfluence}, which indicates that for $F_0 < 0.15$, the field fluence increased during the optimization so that $\overline{F_{\mathrm{opt}}} \approx 0.15$, while for $F_0 \geq 0.15$, $\overline{F_{\mathrm{opt}}} \approx F_0$. A similar result has also been observed for evolution-operator control \cite{BrifGrace2013Talk}.
  
The significant differences between the results of these two sets of simulations confirm that sufficient field strength is necessary for successful optimization. A low initial fluence does not prevent successful optimization if the field strength can increase during the search, as with choice (i) of the control variables.  However, parameterizations of the control that restrict the field strength, such as choice (ii), are only effective for finding optimal fields when the initial fluence is sufficiently large. 

Moreover, choice (ii) appears to lead to a much higher fluence requirement for successful optimization in comparison to the freely varied fields using choice (i).  For choice (ii), a field fluence of $F \approx 2$ was required in order for any searches to succeed.  For choice (i), however, the fluence of many optimized fields was an order of magnitude smaller. This result shows the influence of the parameterization in choice (ii), which introduces constraints beyond those on the field strength.

\subsection{Algorithmic parameters}

Gradient searches may be impeded by algorithmic parameters that prevent accurate solutions to Eq.~\eqref{eq:searchalg}, i.e., severe constraints on the $s$-evolution of the field. This circumstance is especially relevant for search algorithms that employ a fixed step size $\Delta s$.  If the step size is too large, then searches may fail to optimize successfully.  Other constraints on the search algorithm, such as the method used to integrate Eq.~\eqref{eq:searchalg}, may also affect optimization.
 
We performed fixed-step-size gradient optimizations to study how the choice of step size affects the ability to reach a global optimum.  The objective was to maximize $J_P$ for the transition $| 0 \rangle \to | 5 \rangle$ in the system defined in Eq.~\eqref{eq:genhamdip}, with $N = 6$, $\lambda = 1$, and $D = 0.5$. Choice (i) of the control variables was used, and the initial field fluence was $F_0 = 10$. The final time was $T = 50$ and the control period was discretized into $L = 511$ intervals. The objective was optimized with fourth-order Runge-Kutta and forward Euler integrators, and both used a fixed step size chosen on the interval $0.01 \leq \Delta s \leq 0.5$. 1000 optimization runs were performed for each $\Delta s$ value.

\begin{figure}[htbp]
\centering
\includegraphics[width=8cm]{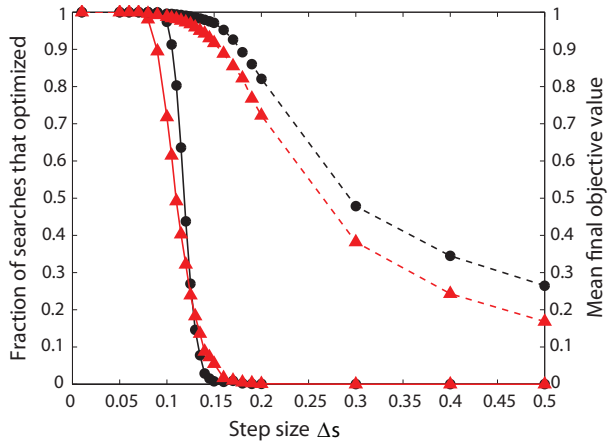}
\caption{(Color online) The fraction of simulations that optimized successfully (solid lines) and the mean final objective value (dashed lines) for fixed-step fourth-order Runge-Kutta (black circles) and Euler (red triangles) methods, as functions of step size $\Delta s$.} 
\label{fig:step}
\end{figure}
Statistical data obtained from these optimizations are shown in Fig.~\ref{fig:step}. With both choices of integrator, at least one search failed for $\Delta s > 0.07$ and every search failed for $\Delta s > 0.2$. In addition, the mean final objective value $\overline{J[\varepsilon(s_f,t)]}$, averaged over the set of 1000 runs, decreased as $\Delta s$ increased. The proportion of searches that failed to optimize for a given $\Delta s$ differs slightly between the two algorithms; the fourth-order Runge-Kutta routine achieves a higher mean objective value than the Euler method for a given step size, but it is computationally slower. For both algorithms, severely constraining the step size will prevent optimization.  

\begin{figure}[htbp]
\centering
\includegraphics[width=8cm]{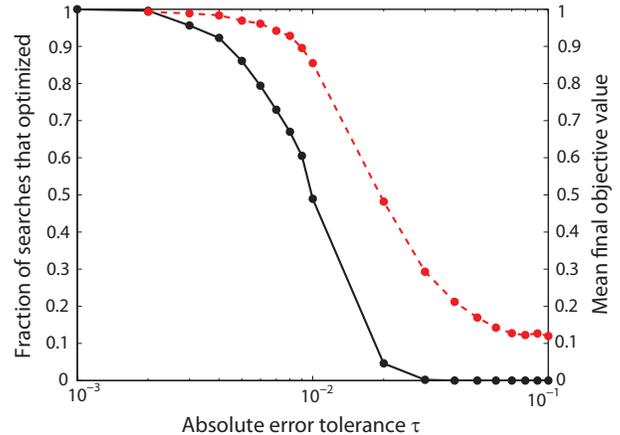}
\caption{(Color online) The fraction of simulations that optimized successfully (solid black line) and the mean final objective value (dashed red line) using \texttt{ode45}, a variable-step-size fourth-order Runge-Kutta method, as a function of error tolerance $\tau$. 1000 optimization runs were performed for each $\tau$ value.} 
\label{fig:error}
\end{figure}

Variable-step routines such as MATLAB's \texttt{ode45} \cite{matlab} estimate an appropriate $\Delta s$ at each step in the search, but this method requires the input of a maximum tolerable error $\tau$ as described in Sec.~\ref{sec:contproc}. This parameter influences the determination of $\Delta s$. We used \texttt{ode45} to perform additional optimizations on the same control problem described above. 1000 simulations were performed for each value of $\tau$ over the range $10^{-3} \leq \tau \leq 10^{-1}$. Figure~\ref{fig:error} shows that at least one search failed to optimize for $\tau > 2 \times 10^{-3}$ and that all searches failed to optimize for $\tau > 2 \times 10^{-2}$. This confirms that excessive error in the solution to Eq.~\eqref{eq:searchalg} constrains the control field and can prevent the achievement of an optimal control.

\subsection{Composite objectives}
\label{sec:compos}

The landscape analysis in Sec.~\ref{sec:structure} applies to the three cost functionals defined in Eqs.~\eqref{eq:pif}-\eqref{eq:w}; collectively, they include the great majority of OCT and OCE objectives. However, some quantum control searches are designed to simultaneously optimize one of these objectives along with one or more other goals. The most common of these other goals is to minimize the field fluence, such that
\begin{equation}
J = J_1 - w \int_0^T \varepsilon(t)^2 dt,
\end{equation} 
where $J_1$ is the primary objective for maximization (e.g., $J_P$ or $J_{\theta}$) and the weight $w > 0$ determines the relative importance of the fluence term. In this case, the results in Sec.~\ref{sec:fieldstrength} suggest that it may be challenging to maximize this objective, since the fluence penalty term constrains the field strength; the significance of this constraint is determined by the value of $w$. It has been shown that such a constraint can prevent the achievement of high values of the primary objective $J_1$ \cite{RenBalintKurti2006, ArtamonovHo2006JCP, ChakrabartiRabitz2007,Kammerlander2011}. More generally, composite objectives involving competitive goals may not exhibit the advantageous landscape structure described in Sec.~\ref{sec:back}. Controls that are critical points of the overall objective $J = \sum_i J_i$, i.e., that satisfy $\delta J/\delta \varepsilon(t) = 0$, are generally \textit{not} critical points of the individual objectives $J_1, J_2, \ldots$ and so it is not possible to simultaneously optimize multiple objectives by including them as terms in a single composite objective. For example, in one numerical study, optimizations of a composite objective relevant to adiabatic quantum computation \cite{BrifGrace2013NJP} encountered local traps. In some OCEs, the cost functional is formulated as a ratio between two objectives (i.e., $J = J_1 / J_2$), and local traps can appear on the corresponding control landscapes of $J$ as well \cite{CardozaTrallero2005, WollenhauptPrakelt2005, BayerWollenhaupt2008}.  Thus, a composite objective may introduce a severe constraint, which can prevent the achievement of a globally optimal value of the individual objective and/or the composite objective, even when other constraints are well-managed.
 
\section{Conclusions}
\label{sec:concl}

The success of quantum control experiments has prompted several works devoted to the theoretical analysis of the landscape critical topology \cite{HoRabitz2006JPPA, WuRabitzHsieh2008JPA, Altafini2009, ChakrabartiRabitz2007, WuPechenRabitz2008JMP, RabitzHsiehRosenthal2004, RabitzHoHsieh2006PRA, RabitzHsiehRosenthal2005PRA, DominyRabitz2008JPA, HsiehHoRabitz2008CP, ShenHsiehRabitz2006JCP}. Collectively, these studies contend that the absence of local optima on the control landscape is responsible for the favorable results in OCEs and OCT simulations. This trap-free topology depends upon three conditions:  controllability, the full rank of the Jacobian matrix $\delta U_T / \delta \varepsilon(t)$, and the unconstrained control field $\varepsilon(t)$. This paper has investigated how gradient-based searches are affected by violating the third condition.

We have shown that the generic favorable properties of the landscape topology can be obscured by placing severe constraints on the control field. We studied the effects of such constraints on OCT searches using a gradient-based algorithm. Artificial traps on the control landscape were observed when the time discretization, number of control variables, control duration, field strength, and algorithmic step size were excessively constrained. These traps are likely to prevent the algorithm from locating a globally optimal control, with the probability of failure typically correlated with the severity of constraint. We have additionally shown that the effect of a constrained parameter on the success of OCT searches may be mediated by other parameters. Importantly, the simulations also demonstrated that no traps are encountered when the constraints are managed properly. Although this paper employs the conservative, myopic gradient algorithm, sufficiently severe constraints can prevent full optimization even with global genetic algorithms.

It has been shown that uncontrollable quantum systems are extremely rare \cite{Altafini2009} and that the presence of singular critical points on the landscape, i.e., the violation of condition (2), appears to produce virtually no risk of trapping in any practically relevant circumstances \cite{RivielloBrif2014}. Combined with these previous conclusions, the present results strongly suggest that the overwhelming majority of encounters with traps ensue from severe control constraints and do not reflect the fundamental landscape character. We conclude that gradient searches performed on controllable quantum systems are extremely unlikely to fail unless the field is severely constrained. Thus, a search that avoids such constraints can take full advantage of the inherently favorable landscape topology.

\acknowledgments

T.-S.H. acknowledges support from the Department of Energy under grant DE-FG02-02ER15344 and H.R. acknowledges support from the Army Research Office under grant W911NF-13-1-0237. R.B.W. acknowledges support from NSFC under Grant Nos. 61374091, 60904034 and 61134008. C.B. was supported by the Laboratory Directed Research and Development program at Sandia National Laboratories. Sandia National Laboratories is a multi-program laboratory managed and operated by Sandia Corporation, a wholly owned subsidiary of Lockheed Martin Corporation, for the U.S. Department of Energy's National Nuclear Security Administration under contract DE-AC04-94AL85000.

\bibliographystyle{apsrev4-1}
\bibliography{riviellorabitz_constraints}

\end{document}